\documentclass[a4,smallextended,final]{svjour3}       
\smartqed  
\usepackage{graphicx}
\usepackage{epstopdf}
\usepackage{dcolumn}
\usepackage{bm}
\usepackage{color}
\usepackage{amssymb, amsmath, bm, textcomp}

\usepackage[T1]{fontenc} 
\usepackage{float}       
\usepackage{mathrsfs}

\usepackage{setspace}    
\usepackage{color}
\usepackage[table]{xcolor}
\usepackage{subfigure}

\newcommand{\dt}{{\Delta t}}

\newcommand{\x}{{\bm x}}

\newcommand{\Kn}{{\mathrm{Kn}}}
\newcommand{\Da}{{\mathrm{Da}}}

\newcommand{\be}{\begin{equation}}
\newcommand{\ee}{\end{equation}}
\newcommand{\bea}{\begin{eqnarray}}
\newcommand{\eea}{\end{eqnarray}}

\usepackage[left=2.1cm,right=2cm,top=3.2cm,bottom=2.8cm]{geometry}

\begin{document}

\title{Mapping Reactive Flow Patterns in Monolithic Nanoporous Catalysts}


\author{Giacomo Falcucci \and Sauro Succi \and Andrea Montessori \and Simone Melchionna \and Pietro Prestininzi \\
Cedric Barroo \and David C. Bell
\and Monika M. Biener \and Juergen Biener \and Branko Zugic \and Efthimios Kaxiras
}


\institute{ G. Falcucci \at Dept. of Enterprise Engineering ``Mario Lucertini'' - University of Rome ``Tor Vergata'',
Via del Politecnico 1, 00100 Rome (Italy) \and John A. Paulson School of Engineering and Applied Sciences - Harvard University, 29 Oxford Street, Cambridge MA 02138 (USA)
\email{giacomo.falcucci@uniroma2.it;        gfalcucci@harvard.seas.edu}\\
\and S. Succi  \at Istituto per le Applicazioni del Calcolo ``Mauro Picone'' - CNR, 
Via dei Taurini, 00159 Rome (Italy) \and John A. Paulson School of Engineering and Applied Sciences - Harvard University, 29 Oxford Street, Cambridge MA 02138 (USA) \\
\and S. Melchionna  \at Istituto per le Applicazioni del Calcolo ``Mauro Picone'' - CNR, 
Via dei Taurini, 00159 Rome (Italy) \\ 
\and A. Montessori \and P. Prestininzi \at Dept. of Engineering - University of Rome ``Roma Tre'', Via della Vasca Navale 79, 00141 Rome (Italy)\\
\and C. Barroo \at John A. Paulson School of Engineering and Applied Sciences - Harvard University, 29 Oxford Street, Cambridge MA 02138 (USA)\\
\and D.C. Bell \at Center for Nanoscale Systems, 11 Oxford St, Cambridge, MA 02138 (USA) \\
\and M.M. Biener \and J. Biener \at Lawrence Livermore National Laboratory, 7000 East Avenue Livermore, CA 94550 (USA) \\
\and B. Zugic \at Department of Chemistry and Chemical Biology, 12 Oxford Street, Cambridge MA 02138 (USA) \\
\and E. Kaxiras \at Department of Physics - Harvard University, 17 Oxford Street, Cambridge MA 02138 (USA) \and John A. Paulson School of Engineering and Applied Sciences - Harvard University, 29 Oxford Street, Cambridge MA 02138 (USA)
}

\date{\normalsize{\textbf{To cite this article}: \\ G. Falcucci et al., Mapping Reactive Flow Patterns in Monolithic Nanoporous Catalysts. \textit{Microfluid. Nanofluid.} 20(7), 1--13 (2016); doi: 10.1007/s10404-016-1767-5; url: http://dx.doi.org/10.1007/s10404-016-1767-5}}

\maketitle

\begin{abstract}
The development of high-efficiency porous catalyst membranes critically depends on our understanding of where the majority of the chemical conversions occur within the porous structure. This requires mapping of chemical reactions and mass transport inside the complex nano-scale architecture of porous catalyst membranes which is a multiscale problem in both the temporal and spatial domain.  
To address this problem, we developed a multi-scale mass transport computational framework based on the Lattice Boltzmann Method (LBM) that allows us to account for catalytic reactions at the gas-solid interface by introducing a new boundary condition. 
In good agreement with experiments, the simulations reveal that most catalytic reactions occur near the gas-flow facing side of the catalyst membrane if chemical reactions are fast compared to mass transport within the porous catalyst membrane. 

\end{abstract}

\section{Introduction}
\label{intro}
\begin{figure}
\centering
	\subfigure[]{\includegraphics[width=0.4\textwidth]{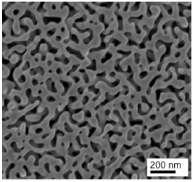}} \;
	\subfigure[]{\includegraphics[width=0.4\textwidth]{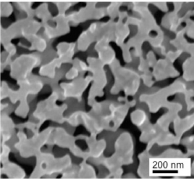}}
	\caption{\label{Fig_1}Evolution of the morphology of np-Au during oxidative coupling of alcohols in a flow reactor. SEM images of the surface of a bulk np-Au sample at the end of the experiment from (a) the surface facing the gas flow and (b) the surface opposite to the gas flow. The surface nano-structure is altered only on the surface facing the incoming gas stream.}
\end{figure}

Catalytic processes lie at the heart of a number of applications relevant to everyday life, including for example fuels, burners, catalytic converters, and fuel cells. For obvious economic, practical and environmental reasons, the optimization of catalytic processes is crucial to many applications. To this end, theoretical and computational studies of molecular interactions at fluid-solid interfaces, where catalytic reactions take place, are central to further optimization of complex real-life catalyst materials. At a fundamental level, this requires a detailed description of bond-breaking phenomena that control electron transfer at the quantum-mechanical level, a process that takes place on the timescale of femtoseconds (10$^{-15}$~s). Tracking these phenomena on temporal and spatial scales of experimental relevance is computationally impossible, hence the need of multiscale techniques, connecting the atomic level to the meso- and macroscopic scales, \cite{Salsiccioli_2011,Keil_2012,Succi_2001,Ma_2015,MacMinn_2015,Bording_2001}.
Multiscale computational models capable of describing reactive flows in monolithic catalytic media are urgently needed to assess the catalytic activity and stability of new catalyst materials such as the recently discovered class of monolithic nanoporous gold 
(np-Au) \cite{Wittstock_2010,Wittstock_2014,Parida_2006}. Despite the generally high stability of np-Au, reaction induced coarsening of both pores and ligaments has recently been observed during oxidative coupling of alcohols under flow reactor conditions, \cite{Stowers_2015}. The coarsening seems to depend strongly on the orientation of the sample with respect to the gas stream, with substantial coarsening of the rear surface while the gas-facing side appears to be unaltered (see Fig. \ref{Fig_1}). 

This behaviour suggests substantially different local gas-surface chemistries on the up- and down-stream catalyst surface. Here, multiscale computational models predicting the reactive flow through a monolithic catalytic sample can provide critical information about the local gas-surface chemistry and about the locations of  reactions within the catalyst.    
Here, we report on the development of a predictive multi-scale mass transport computational framework based on the Lattice Boltzmann Method (LBM), \cite{Succi_book,BSV}, that allows further efficiency improvements by guiding the development of catalyst architectures. As a first step toward such a multiscale description, we develop a model for mesoscale reactive flows in porous media that takes into consideration the nano- and microscale  structure of the pores. From a macroscopic point of view, the physics of fluid-solid interactions is captured by appropriate boundary conditions, which specify the amount of mass, momentum and energy exchange between impinging and desorbing molecules. The microscopic behaviour is then analysed to yield effective reflection/absorption coefficients, \cite{Sbragaglia_2005}. LBM is a mesoscale technique based on a minimal (lattice) version of Boltzmann's kinetic equation, that has proven to be a fast and reliable numerical tool for the investigation of nano- and meso-scale phenomena, especially in the presence of non-trivial boundary configurations, \cite{Succi_book,BSV,Fyta_2006,Bernaschi_2010,Bernaschi_2013}. 
It has been successfully employed in the last two decades for the simulation of complex fluid dynamics phenomena, \cite{Aidun_2010,Biscarini1}, such as blood circulation \cite{Bernaschi_2009}, multiphase flows \cite{Shan_1993,Swift_1995,Falcucci_2007}, cavitation \cite{Falcucci_2013}, and fluid-structure interaction \cite{LBM_FSI}. Flow in porous media with heterogeneous catalysis is yet another very active area of LBM research, \cite{Montessori2016,Biscarini2,Biscarini3,Succi_2002,Succi_2002_PRL}. 
In the present implementation of LBM, we introduce a new boundary condition to account for chemical reactions and transport phenomena inside nano-scale pores of monolithic porous catalyst membranes. The conversion efficiency predicted by our model is compared to the experimentally observed conversion rates for methanol oxidation over a nanoporous gold disk using a quartz tube microreactor. This catalyst has recently been reported to be highly effective for the selective oxidation of methanol to methyl formate, \cite{Wittstock_2010_2,Wittstock_2010_3,Kosuda,Stowers,Wang,Personick,Fujita}. In general, we observe good agreement in terms of conversion efficiency and field distribution of reactant ($R$) and product ($P$) species thus demonstrating that our model provides a powerful tool for future design optimization of nanoporous catalyst architectures.

\section{Methods}

\subsection{Experimental details and Nanoporous Catalyst reconstruction}
\begin{figure}[h!]
\centering
	\includegraphics[width=0.5\textwidth]{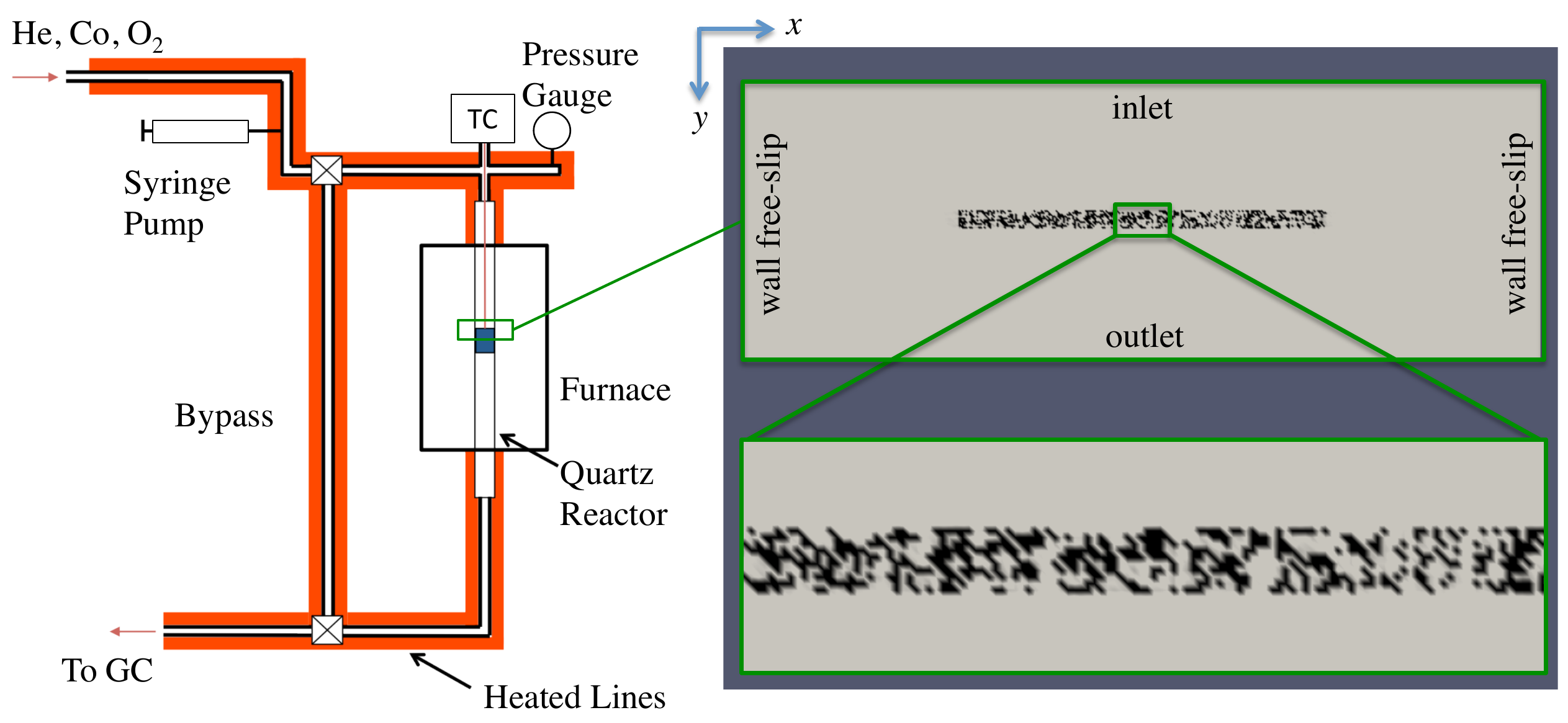} 
	\caption{\label{Fig_7}Schematic representation of the experimental flow reactor (left panel) and the simulation domain (right panel). The latter consists of a grid of 500$\times$250 lattice units (representing a 5 $\mu$m $\times$ 2.5 $\mu$m section of a 2D reactor with 10 nm resolution). The porous slab (ingot), with a typical aspect ratio of $w/h = 0.05$ and a typical pore diameter of $\sim$30 nm ($\sim$ 3 lattice units) is centred in the middle of the domain. }
\end{figure}

Nanoporous gold catalyst samples for the methanol oxidation reaction were prepared by dealloying bulk Ag$_{70}$Au$_{30}$ alloy discs \cite{Biener_NanoLett}, (5 mm diameter, 200-300 $\mu$m thick) in $70\%$ nitric acid (Alfa Aesar) for 48 hours. After washing thoroughly in deionized water and drying in static air, the monolithic catalyst was loaded into a quartz 
tube microreactor for catalytic testing. Prior to use, the catalyst was activated according to a recently developed procedure, \cite{Gordon,Rohe}. A schematic representation of the reactor system used is shown in the left panel of Fig. \ref{Fig_7}. 

Selective methanol oxidation was used as a test reaction:
\begin{equation}
2 \ \text{CH}_3 \text{OH} + \text{O}_2 \rightarrow \text{HCOOCH}_3  + 2 \ \text{H}_2 \text{O}
\end{equation}
The reaction was carried out at 150$^{\circ}$C in a continuous flow of reactant (6.5$\%$MeOH - 20$\% $O$_2$ - 73.5$\%$ He) at 50 ml/min. The effluent gas was analysed using a gas chromatograph (Agilent HP 7890A) coupled to a mass spectrometer (MSD 5975C), and equipped with two columns operated in tandem (HP-PLOT/Q and CARBONPLOT). Scanning electron microscopy (SEM) analysis was performed on a Zeiss Supra 55VP instrument. 
The simulation domain consists of a grid of 500$\times$250 lattice units (representing a 5 $\mu$m $\times$ 2.5 $\mu$m section of a 2D reactor with 10 nm resolution). The porous medium  (representing np-Au) was constructed by growing a number of $N$ disks with 30 nm diameter (representing pores) out of randomly chosen centres within a rectangular area centred in the middle of the simulation domain. The number of disks was increased until the porosity $\phi$ (void volume/total volume) within the rectangular area reached a value $\phi \sim$  0.70 $\pm$0.005, the typical void fraction of np-Au made from Ag$_{70}$Au$_{30}$ ingots. The value of the porosity is fixed throughout the simulation, since we assume that the structure of the catalyst ingot does not change in time.\\
The relatively small computational cell permits us to match the most important experimental parameters, such as the typical diameter-to-thickness aspect ratio of np-Au ($\sim$20), and allows us to resolve 30 nm pores, but restricts the sample size and pore length-to-diameter aspect ratio that can be reproduced by our simulations. The simulation domain and the porous medium are shown in the right panel of Fig. \ref{Fig_7}. 
For experimental visualization of the gas diffusion kinetics in np-Au, we performed alumina (Al$_2$O$_3$) atomic layer deposition (ALD) experiments. Rather than using long saturation exposures that result in formation of homogeneous coatings \cite{Ott,Biener}, here we worked in the diffusion limited regime by using short trimethyl-aluminum (TMA)/H$_2$O ALD cycles (1s/1s at 0.8 torr, 125$^{\circ}$C). The cross-sectional distribution of the deposited alumina, reflecting the diffusion kinetics of TMA in np-Au, was determined by cross-sectional SEM analysis of the fractured np-Au sample.

\subsection{Numerical Model}
\begin{figure}
\begin{center}
	\includegraphics[width=0.5\textwidth]{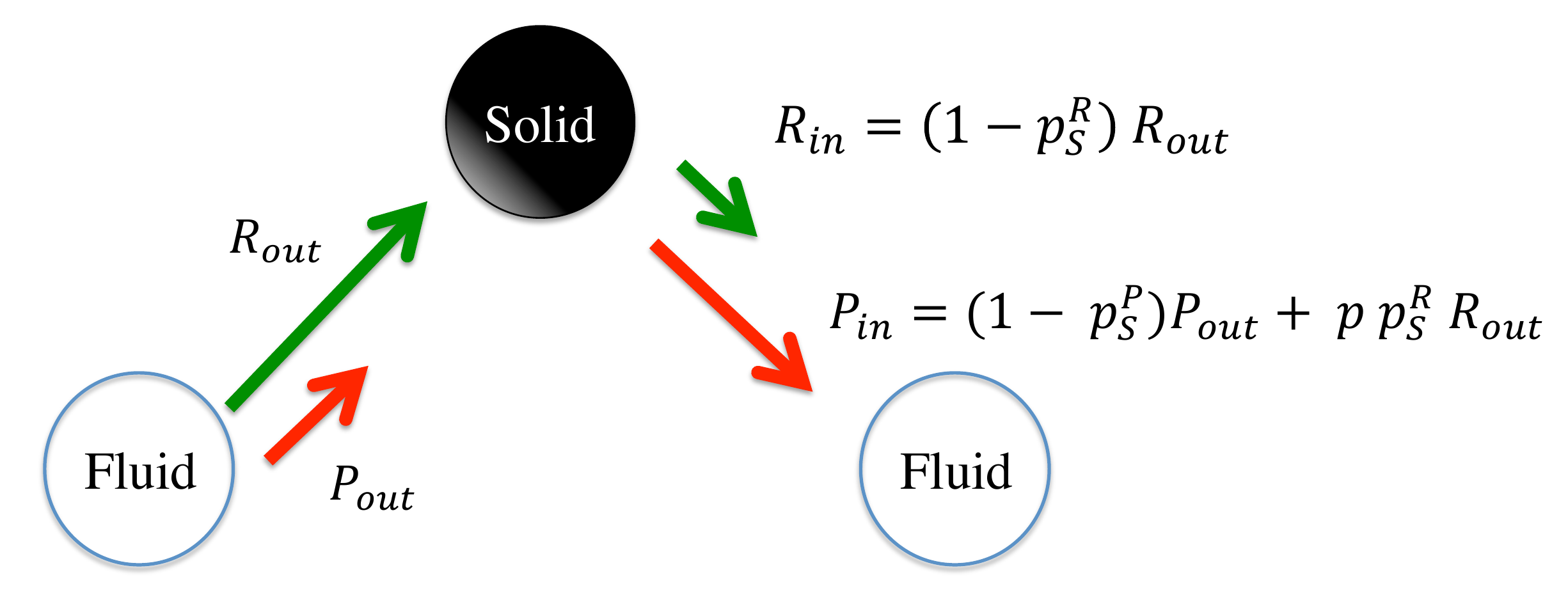} 
\end{center}
	\caption{\label{Fig_8}Sketch of the Stick and React boundary model (see text).}
\end{figure}

The basic idea behind LBM is to reproduce the microscopic dynamics of groups of particles in the Eulerian framework, \cite{Succi_book,BSV}. According to the kinetic picture of statistical mechanics, the microscopic evolution is described by the probability of finding a molecule, at given time $t$, at the position $\bm{x}$ with velocity $\bm{c}$. The interaction between particles is then described as a sequence of collisional events between the gas molecules in the bulk and gas-solid molecules at the interfaces. In the continuum flow regime, gas-gas collisions are orders of magnitude more frequent than gas-solid interactions. For diffusion through a nanoporous solid, when the mean free path between gas-gas collisions becomes comparable to the pore diameter, the two processes take place with approximately the same frequency.
The Boltzmann equation lives in a six dimensional phase space (ordinary space plus velocity space), which is very hard to handle numerically. Because of this, LBM uses unconventional ways to reduce the computational effort without losing physical accuracy. The main features that underlie the efficiency of the method are reported in the Supplementary Material Section. Here, we briefly recall the reasons why a mesoscale model like LBM, equipped  with suitable boundary conditions, may achieve an optimal compromise between physical fidelity and computational viability. LBM is known for its versatility to scale both ways, upwards to continuum fluid mechanics and downwards to molecular scales. In this work, we have leveraged both, anchoring the resolution to the nanoscopic level (grid spacing 10 nm). LBM has proven to yield reliable quantitative information also in the finite-Knudsen regime, once equipped with proper kinetic boundary conditions that capture fluid-wall mass/momentum transfer, \cite{Ansumali,Succi_2002,Montessori_2015,Niu_2007}. 
In our model, such boundary conditions are enriched with catalytic reactions occurring between fluid/gas molecules and solid bulk: upon colliding with a solid site, the reactant $R$ partially sticks to the wall with probability $p_S^R$  (see Fig. \ref{Fig_8}). This fraction reacts with probability $p$, developing products at the pore surface. These product species then re-enter the pore region with probability $(1-p_S^P)$, where $p_S^P$ is the sticking probability of the products. Non-converted reactant molecule re-enter the pore with probability $(1-p_S^R)$.

The physics under investigation could be described by solving the full Boltzmann equation using Direct Simulation Monte Carlo techniques (DSMC), \cite{Bird}.  In practice, however, DSMC would be computationally unfeasible, due to the large statistical fluctuations, which are known to scale like the inverse of squared Mach number. Given the extremely small values of the Mach number in the present application ($\mathrm{Ma}\sim~10^{-3}$), the use of DSMC is not suited to our purposes \cite{DiStaso}.

\subsection{Dimensionless Numbers}
We report here the values of the main dimensionless parameters controlling both the evolution of the carrier and the chemical behaviour of reactants and products species, specifically the Reynolds ($\mathrm{Re}$), Knudsen ($\mathrm{Kn}$), Pecl\'et ($\mathrm{Pe}$) and Damk\"ohler ($\mathrm{Da}$) numbers. The Reynolds number expresses the ratio between inertial and viscous forces:
\begin{equation}
\mathrm{Re}= U_{in} \ h /\nu           
\end{equation}
where $U_{in}$ is the gas inflow velocity, $h$ is the ingot diameter and $\nu$ is the kinematic viscosity of the gas.
In our simulations,  $U_{in}=0.08$  $lu/\Delta t$,  $h=200$ $lu$, $\nu=7/3$ $lu^2/\Delta t$ ($lu$ represents the grid spacing and $\Delta t$ is the lattice time step) providing $\mathrm{Re}_{LBM} \sim 10$, which is in good agreement with the experimental one ($\mathrm{Re}_{exp} = 10$).
The Knudsen number  is the ratio between the molecular mean free path and the characteristic pore diameter 
\begin{equation}
\mathrm{Kn}=c_s (\tau- \Delta t/2) \ 1/\delta
\label{knudsen}
\end{equation}
where $\delta$ is the typical pore size ($\sim$ 3 in lattice units), $\tau$ is the collision time.
Equation \ref{knudsen} stems from the expression of the mean free path in LBM fluids, 
$\lambda = c_s \left( \tau - \frac{\Delta t}{2} \right)$.  
In the present simulations, $\tau \sim  4 \ \Delta t$, $\delta \sim 3$ and the lattice speed of sound $c_s=1/\sqrt{3}$, yielding $\mathrm{Kn} \sim 0.6$, marginally in the Knudsen diffusion regime. Since the experimental value is $\mathrm{Kn}\sim 2$, our numerical model is expected to capture the basic non-equilibrium effects due to flow rarefaction within the nanoporous gold sample in the transitional regime ($0.1 \leq \mathrm{Kn}  \leq 5$).  The chemical reactivity is controlled by two non-dimensional numbers, namely, the Damk\"ohler number, $\mathrm{Da}$, expressing the ratio between the characteristic chemical and transport time scales, and the Pecl\'et number, that is the ratio between convection and diffusion:
\begin{eqnarray}
\mathrm{Da}  &=& \tau_{ch}/(\delta / v_{th}) \label{Da_def} \\
\mathrm{Pe}  &=& U_{in} \cdot h/D  \label{Pe_def}         
\end{eqnarray}
where $D$ is the diffusion coefficient of the gas and $v_{th} = \sqrt{\frac{k_B \ T}{m}}$  is the gas thermal speed. \\
The characteristic time of the reaction $\tau_{ch}$ is a function of the time step expressed in physical units, $\Delta t_{exp}$, and the reaction probability $p$, $\tau_{ch}~=~\Delta t_{exp} / p$. Taking $p$ as 0.45 and retrieving $\Delta t_{exp}$ form the compliance between experimental and numerical viscosities ($\Delta t_{exp} = \nu_{LB} \ \Delta x^2 / \nu_{exp}$), we have
\begin{equation}
\tau_{ch} \sim 25 \ \text{ps ,}
\label{tau_ch_num}
\end{equation}
providing Damk\"ohler number $\mathrm{Da}  \sim 0.4$. Finally, we find $ \mathrm{Pe} \sim 8$, which is very close to the experimental value of Pecl\'et number, $\mathrm{Pe}_{exp}=10$.
Table \ref{nondimpar} summarizes the main dimensionless parameters of our simulations.\\
\begin{table}[h!]
\begin{center}
\begin{tabular}{l l l l l}
	$\mathrm{Kn}$ & \quad $\mathrm{Pe}$ & \quad $\mathrm{Da}$ & \quad  $\mathrm{Re}$ & \quad  $\mathrm{Ma}$ \\
\hline
   $0.6 $  				&  \quad  $8.0$				&    \quad $0.4$				& \quad  $10$				& \quad  $10^{-3} $  \\
\hline
\end{tabular}
\end{center}
\caption{\small{\label{nondimpar}Main dimensionless parameters of the simulations.}}
\end{table}

\subsection{Consistency of Reaction Time}
A key element of our model in terms of comparison to experiment is the relative rate of reaction on the surface as compared to diffusion rate throughout the nanoporous structure; the latter is captured by the values of $\mathrm{Re}$, $\mathrm{Kn}$ and $\mathrm{Pe}$. The former is taken into account by the value of $\mathrm{Da}$, which depends on the chemical reaction time $\tau_{ch}$. We provide an independent check of the chemical time scale value $\tau_{ch}$, as given in Eq. (\ref{tau_ch_num}), by comparing with the time scale computed from ab-initio Density Functional Theory (DFT) calculations, \cite{Xu_2011}. We define the reaction rate (inverse time scale) as:
\begin{equation}
R=N_c \  A       
\end{equation}
where $N_c$   is number of catalyst atoms in the lattice cell  (lattice cell is 10 nm $\times$ 10 nm in our simulation), $A$ the Arrhenius factor  $A=k \ e^{(-(E/(k_B T)) )}$, with  $k$  the attempt rate (1/s). Given that the catalyst density is
\begin{equation}
n_c=\frac{10 \ atoms}{(10 \ \text{nm})^2}
\end{equation}
we estimate $N_c=10$ atoms in a LBM cell; taking $ k=10^{13} \ \text{s}^{-1}$, $E=0.35$ eV and $T=450$ K (see \cite{DiStaso}), which are typical values for the reactions considered here, we obtain an effective reaction rate:
\begin{equation}
R \sim 1.0 \ \times \ 10^{10} \   \text{s}^{-1}
\end{equation}
As a result, the characteristic chemical time scale is equal to:
\begin{equation}
\tau_{ch}=1/R \sim \ 100 \ \text{ps}  
\label{tau_ch_exp}     
\end{equation}
that is, the same order of magnitude of $\tau_{ch}$ in Eq. (\ref{tau_ch_num}), computed according to the simulation time step.  Considering the experimental value of the volumetric flow  50 mL/min   corresponding to about  $\dot{N} =  3 \times 10^{19}$  Reactant molecules/s,  and an effective reaction probability \cite{Wang_2015}  $p_{eff}=10^{-11}$ obtained from experimental measurements, we find  $\dot{R}  = \dot{N} \cdot p_{eff}  \sim 3 \times 10^8$~reactions/s in the whole ingot. \\ 
We can further define the chemical time scale as $\tau_{ch}=(1/R)/(S/A)$, which is the ratio of the exposed surface area within the pores versus the area of the ingot gas-facing area. Based on digital reconstruction of the nanoporous material, we obtain $S/A \sim 200$ for a reactive ingot layer of about 10 $\mu$m in depth, which gives $\tau_{ch} \sim 20$ ps, in close match with the values of $\tau_{ch}$ from LBM evaluation and DFT prediction.

\section{Results \& Discussion}
Table \ref{Tab_1} reports the main physical parameters of our baseline simulations, in lattice non-dimensional units (second column) and in physical units (right column), respectively.
\begin{table}
\begin{center}
\begin{tabular}{l c l}
\hline
	& Lattice units & Physical Units \\
\hline
$L$	 				& 		250			&		$2.5 \ \times \   10^{-6}$  m \\
$H$					&		500			&		$5 \ \times \   10^{-6}$  m \\
$h$					&		200			&		$2 \ \times \   10^{-6}$  m \\
$w$					&		10				&		$10^{-7}$ m \\
$\rho_{inlet}$	&		1				&		$0.833$  mL/s \\
$\delta$			&		3				&		$3 \ \times \   10^{-8}$  m	 \\
$\bar{v}$			&	  $1/\sqrt{3}$	&		$\sim 1000$ m/s \\
$\tau_{tr}$		&  3 $\sqrt{3}$	&	$\sim 5  \ \times 10^{-11}$ s \\
$\tau_{ch}$		&	2	~				& 		$2.5 \ \times \  10^{-11}$ s \\
$p_S^R$			&		0.95			&		NA \\
$p_S^P$			&		0.00			&		NA \\
$p$					&		0.45			& 		NA \\
\hline
\end{tabular}
\end{center}
\caption{\small{\label{Tab_1}Main physical and chemical parameters adopted in our simulations, with the chosen values in lattice non-dimensional units and the corresponding parameters in physical units; the parameters are computed for the reference case, corresponding to $w = 10$ and $p = 0.45$.}}
\end{table} \\

\begin{figure}
\centering
	\includegraphics[width=0.44\textwidth]{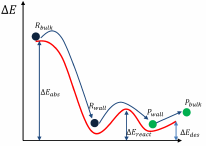} 
	\caption{\label{Fig_2}Sketch of the energetic landscape associated with absorption of reactants within the surface and desorption of reaction products back into the pore. Once they come sufficiently close to the surface, the reactant are attracted to it with an absorption energy of $\Delta E_{abs} \sim~0.8 - 1$ eV. Subsequently, they turn into products by going over an activation barrier of $\Delta E_{react} \sim~0.4$ eV. The reactant face with a much smaller desorption barrier of $\Delta E_{des} \sim~0.1 - 0.2$ eV.}
\end{figure}

In non-dimensional lattice units, the initial values for the densities of the carrier ($\rho_C$), and the two reactive species 
($\rho_R$  and $\rho_P$) inside the computational domain, are set as follows:
\begin{equation}
\rho_C=0.8, \quad   \rho_R=\rho_P=0                     
\end{equation}
Neither reactant nor product species are present at the beginning of the simulation. At the inlet, $x=0$, carrier and reactant are injected at a constant speed, $U_{in}=0.08$, density $\rho_C=0.8$ and $\rho_R=0.2$. Once the reactant reaches the porous slab, products start to appear according to the chemical reaction $R \rightarrow P$. Both $R$ and $P$ species are passively transported by the carrier and experience Knudsen diffusion within the pores. 
The sticking probabilities of the reactant and products are kept fixed to $p_S^R \sim 0.5$ and $p_S^P \sim 0.0$, respectively.
These values have been chosen based on input from lower-scale electronic structure simulations, which give absorption energies of the order of $0.8-1.0$ eV and desorption barriers of about  $0.1-0.2$ eV. These values clearly indicate that reactants near the surface  are attracted and stick to it,  
while the products experience with a much smaller desorption  barrier, hence their desorption from the surface is highly likely. The ratio between the corresponding Arrhenius rates at 150$^{\circ}$ Celsius degrees is of the order of $10^{-6}$, which fully justifies the choice of a zero sticking coefficient for the products. The schematic illustration is reported in Fig. \ref{Fig_2}.

\begin{figure*}
\centering
	\subfigure[]{\includegraphics[width=0.49\textwidth]{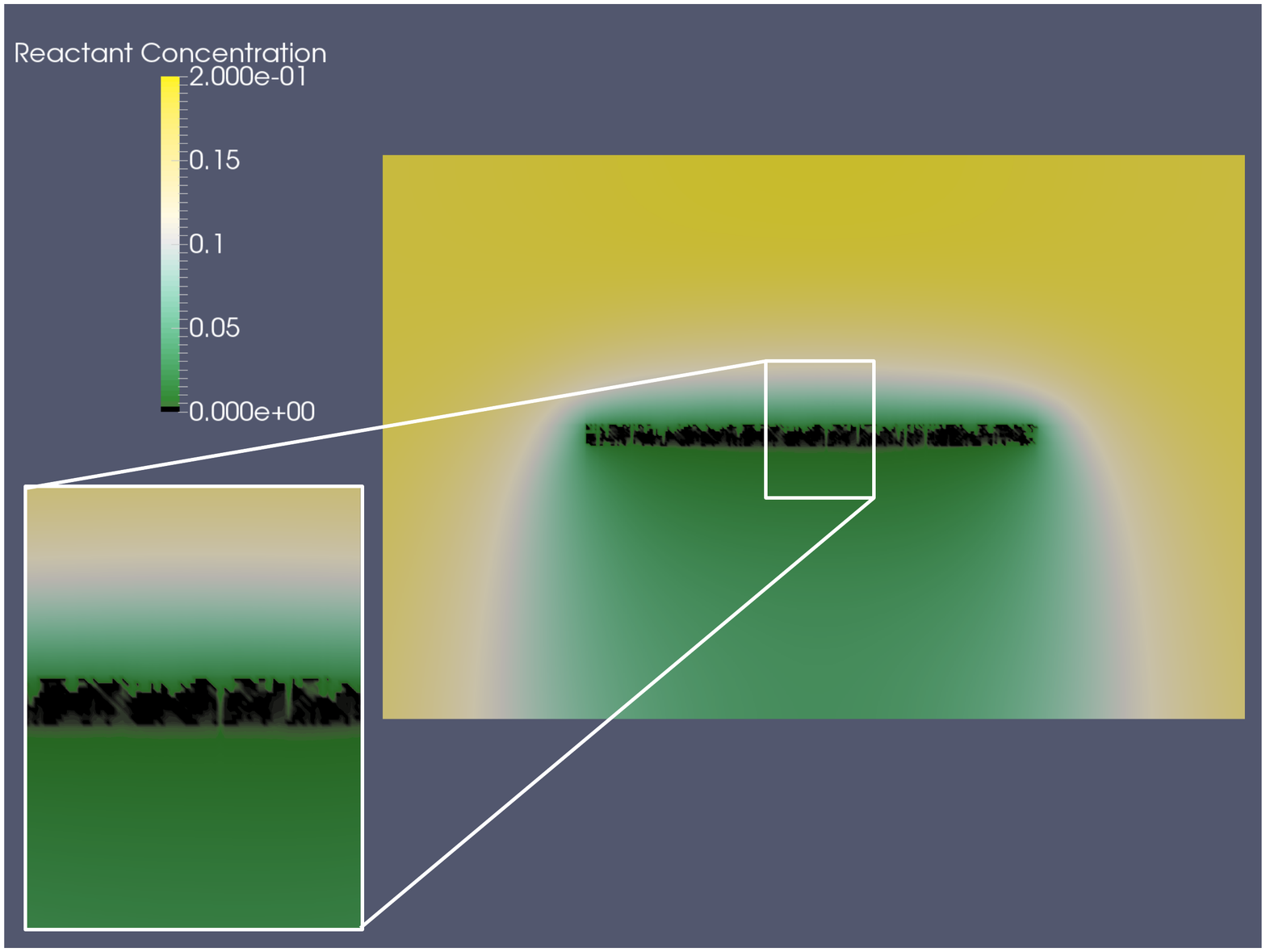}} \,
	\subfigure[]{\includegraphics[width=0.49\textwidth]{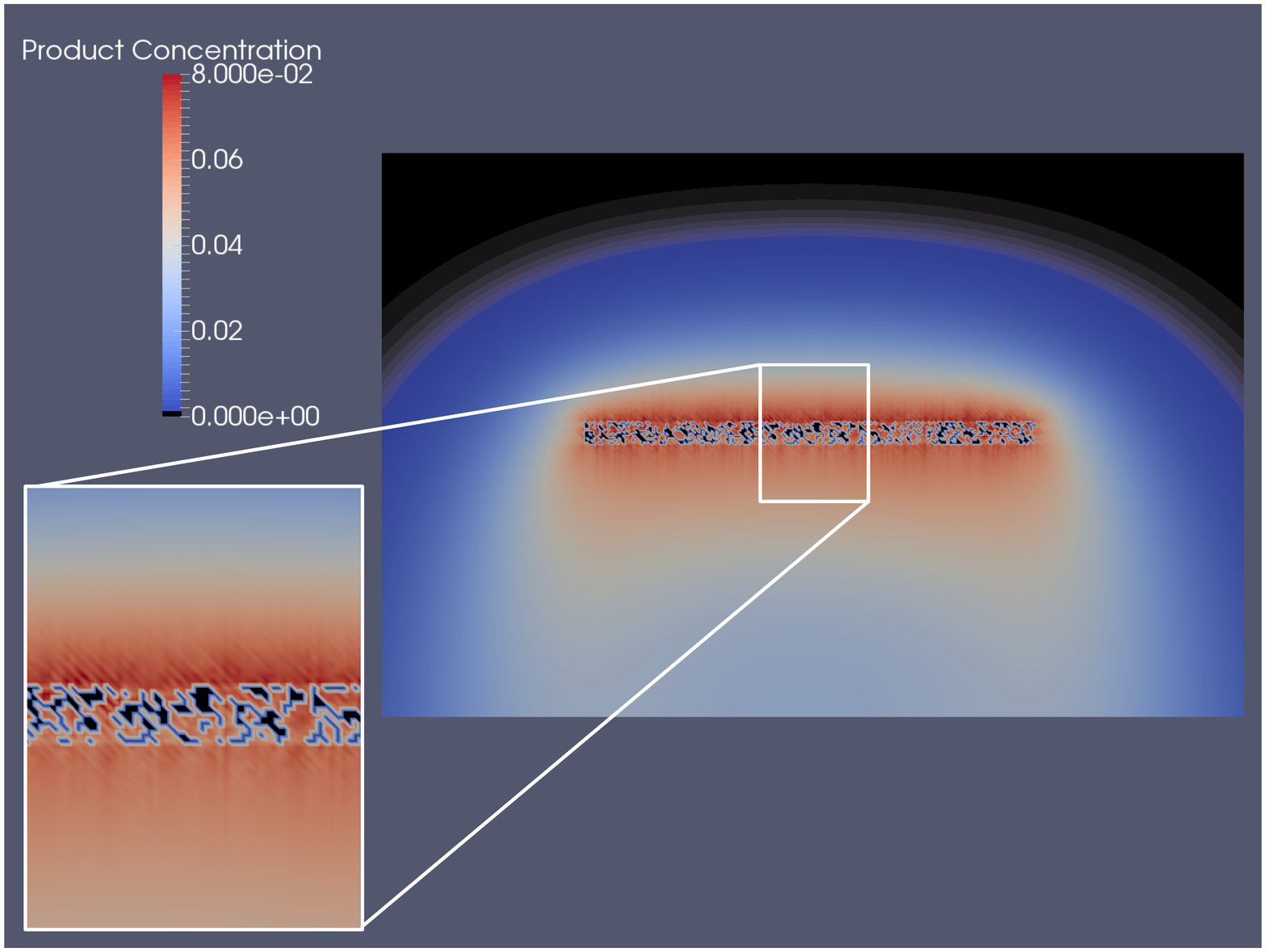}} \\
	\subfigure[]{\includegraphics[width=0.49\textwidth]{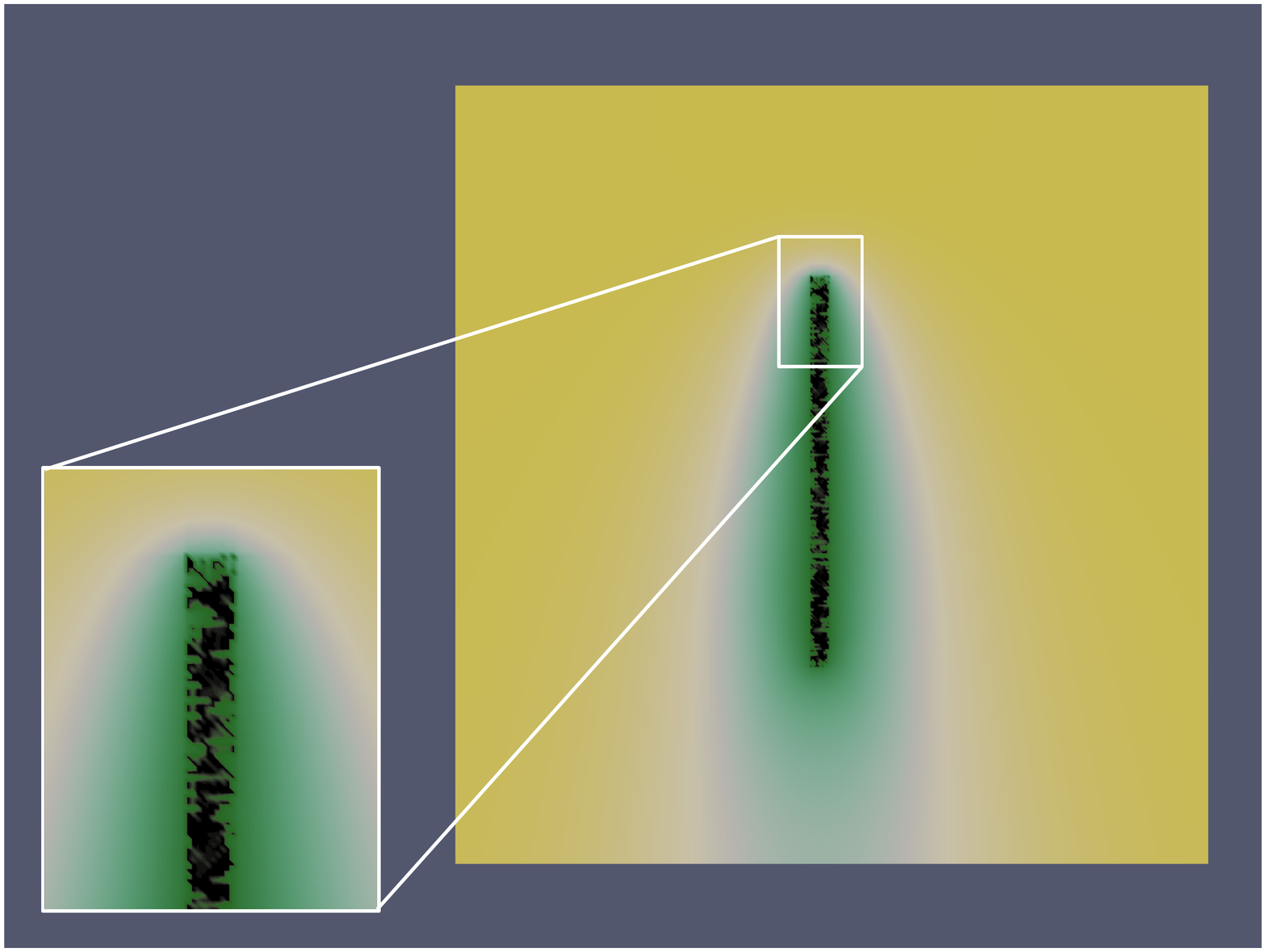}} \,
	\subfigure[]{\includegraphics[width=0.49\textwidth]{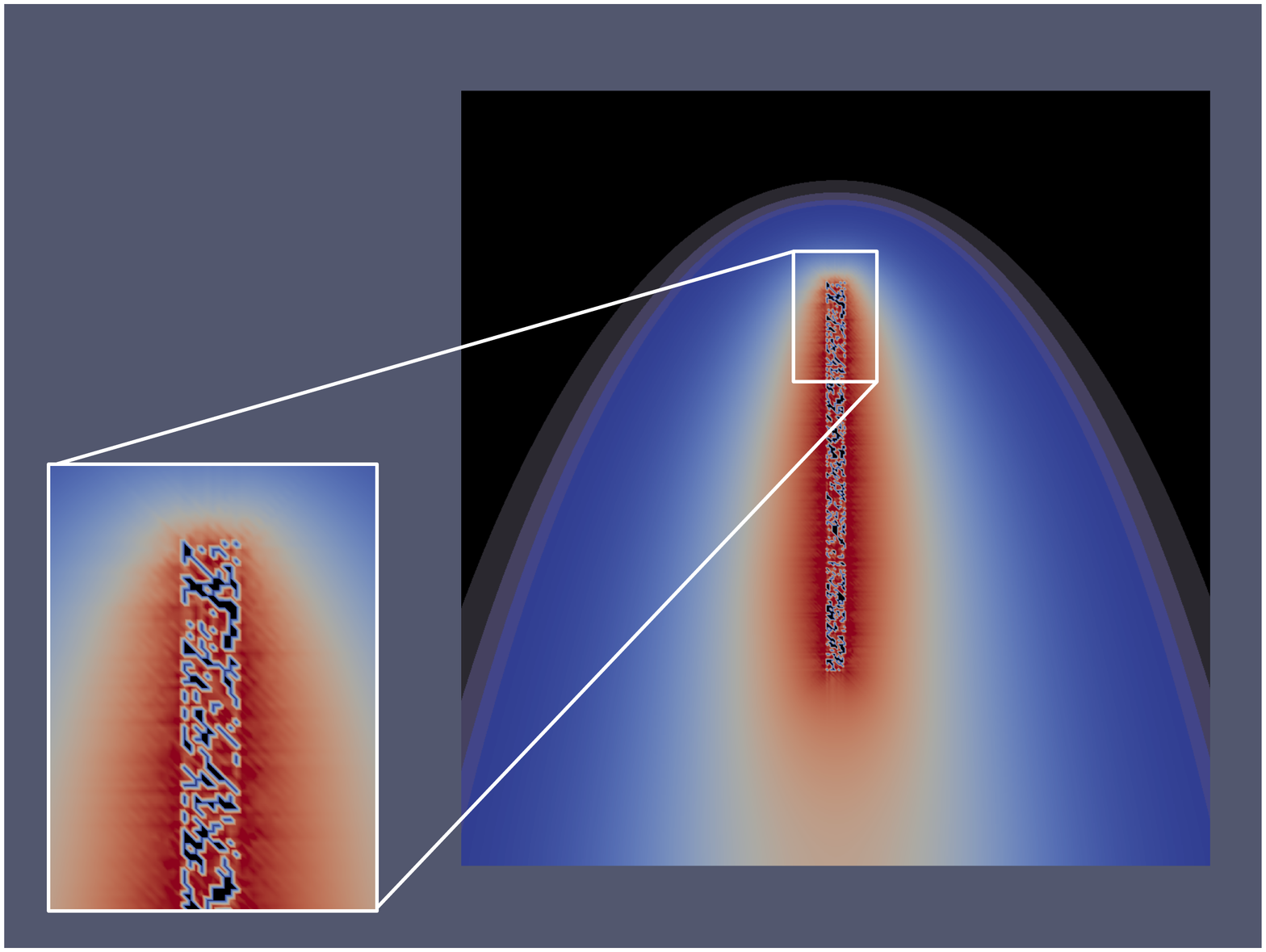}} 
	\caption{\label{Fig_3}Examples of reactant and product concentrations at $t = 10000$ time steps, for the value of the reaction parameter $p = 0.45$,  for perpendicular [(a) and (b)] and parallel [(c) and (d)] orientations of the long axis of the ingot relative to the flow direction, which here is from top to bottom in each figure.}
\end{figure*}

Our prime goal was to systematically investigate the effect of the reaction probability and the ingot geometry (thickness $w$ and ingot diameter $h$, corresponding to the sample thickness and diameter in 3D) and of the sample orientation on conversion efficiency and product distribution.  For 2D simulations, in the presence of bulk conversion, the conversion efficiency $\eta$ is expected to scale with the area $A = h \ w$, (corresponding to the volume $V = h^2 \ w$ of a 3D ingot), while for reactions taking place only at the ingot outer surface normal to the gas flow (that is, the upper side of the ingot in Fig. \ref{Fig_3}, the rear (lower) side of the ingot being almost inactive), 
$\eta$ would scale like $p_u=h+2w$. 

Finally, for full perimeter conversion, the efficiency would scale with the total perimeter $p_{tot}=2h+2w$. These scaling relationships rely on an assumption of homogeneity, which is eventually broken at the corners of the ingot, where larger gradients are found. Note that the scaling is the same as for a 3D slab of cross-section $h \times h$ and depth $w$, with $w << h$. 
The essence of the simulations is captured by the density contour plots of the reactants and products as they flow around and through the catalyst ingot. Representative results are shown in Fig. \ref{Fig_3}, for two possible orientations of the ingot relative to reactant flow direction, perpendicular to the flow (horizontal ingot) and parallel to the flow (vertical ingot) direction.  Inspection of the (steady-state) density contours of the reactant $\rho_R (x,y)$ and product densities $\rho_P (x,y)$ reveals that: 
\begin{enumerate}
\item the reactants do not diffuse very deep into the porous structure, as they are efficiently converted to product molecules on the gas-facing surface of the ingot; 
\item the majority of the products leave the porous structure through the inlet-facing surface, while a smaller fraction diffuses through the sample and leaves through the back face. 
\end{enumerate}
As a consequence  of the efficient conversion of the reactant molecules to products within a shallow catalyst layer on the side facing the gas stream, the rear face of the ingot appears to be shadowed and does not contribute substantially to the conversion efficiency (see Figs. \ref{Fig_1} and \ref{Fig_3}). 
\begin{figure}
\centering
	\includegraphics[width=0.44\textwidth]{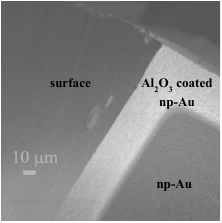} 
	\caption{\label{Fig_4}Cross-sectional SEM image of a fractured, $200$ $\mu$m thick np-Au sample after exposure to 1 TMA/H$_2$O ALD cycle (1s/1s at 0.8 torr). The 1 s long sequential TMA/H$_2$O exposure results in the deposition of a one monolayer thick Al$_2$O$_3$ coating on the ligaments of np-Au within a $\sim 20$ $\mu$m thick surface layer (bright area of the cross section) separated by a sharp interface from uncoated np-Au.}
\end{figure}

\begin{figure}
\centering
	\subfigure[]{\includegraphics[width=0.48\textwidth]{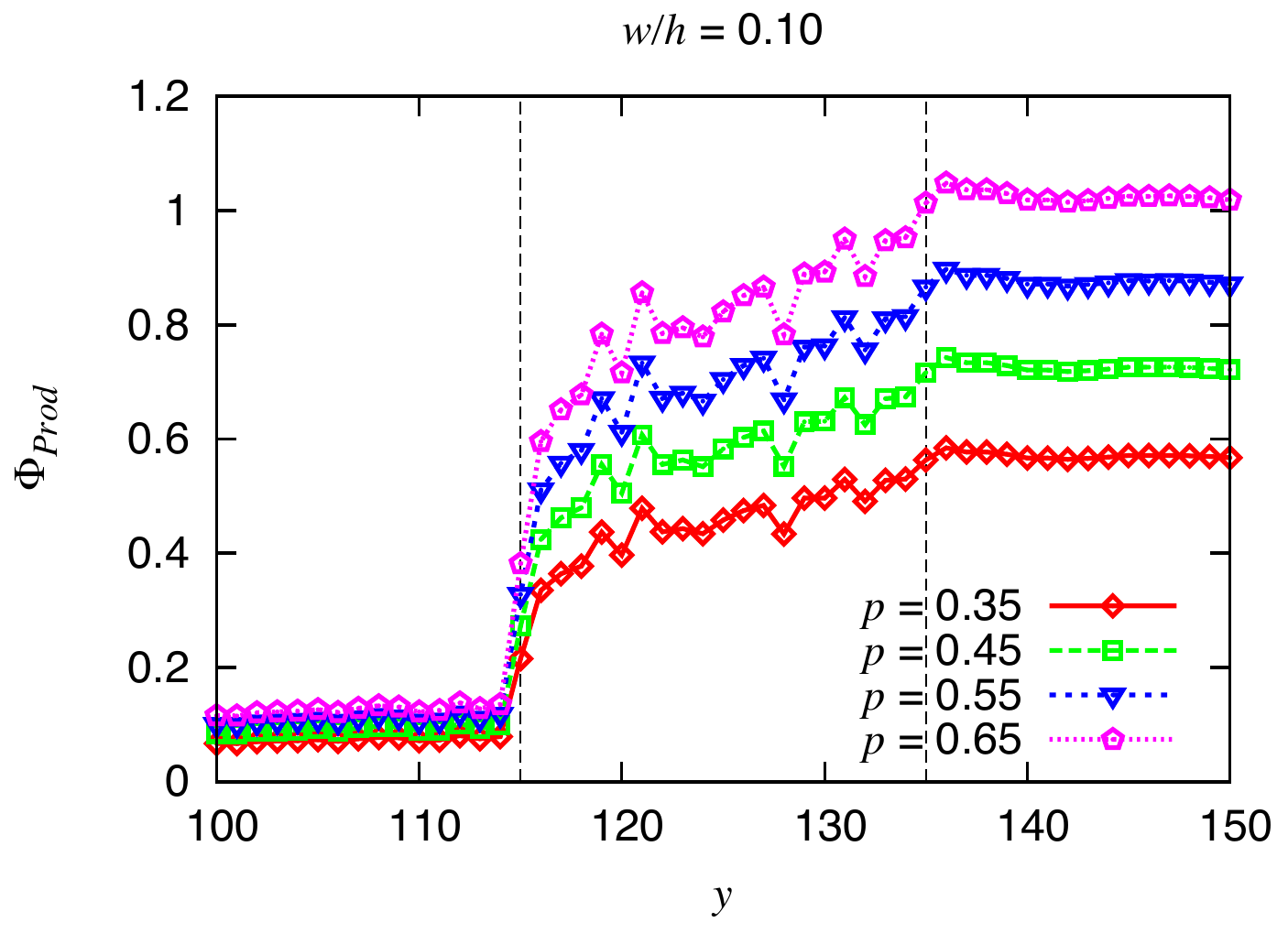}} \qquad
	 \subfigure[]{\includegraphics[width=0.48\textwidth]{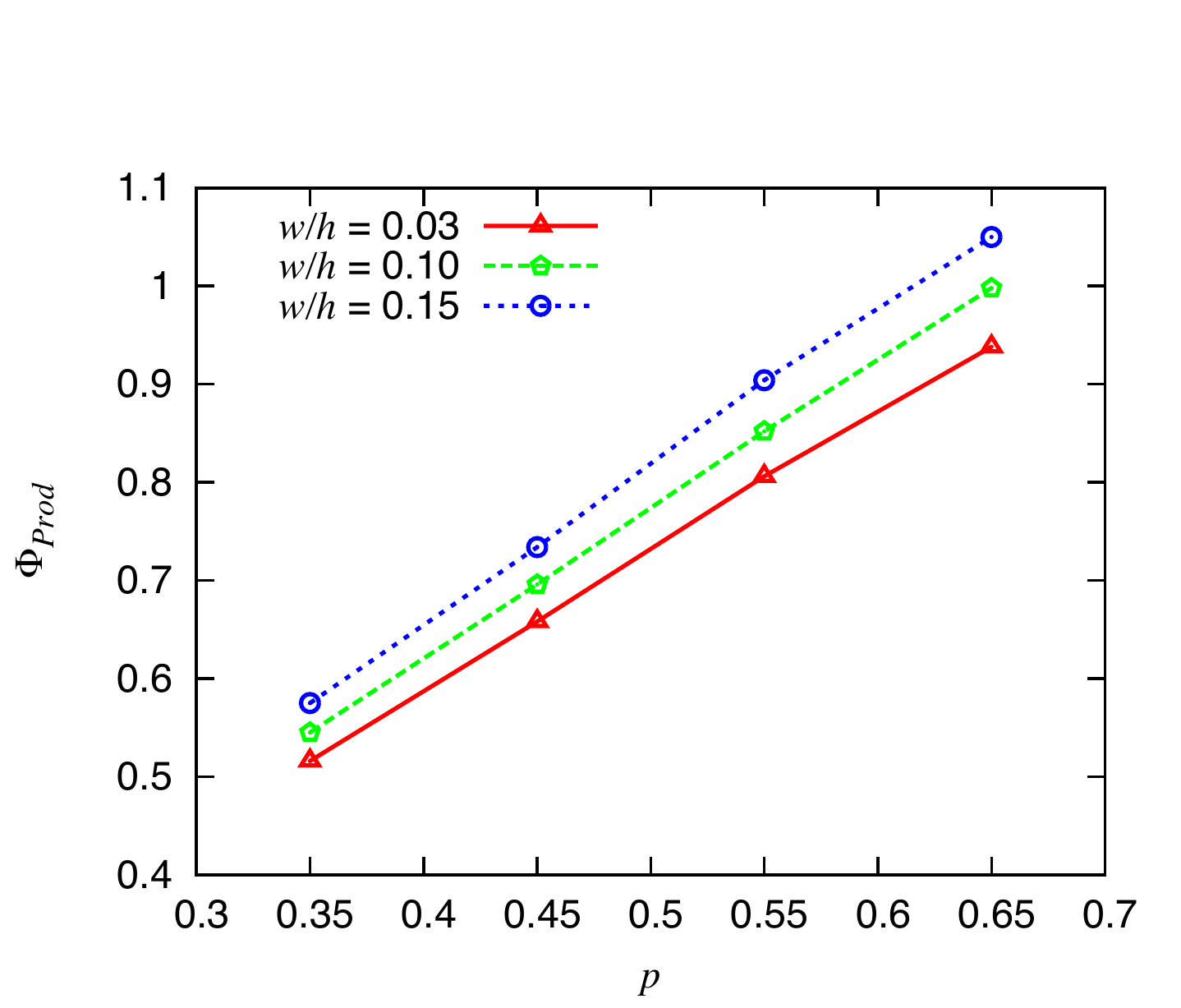}}
	\caption{\label{Fig_5}: Left panel: Flux of the product species along the main flow direction (top to bottom in Figure \ref{Fig_3}) for different values of the reaction probability $p$ and $w/h = 0.1$, after $t=30000$ time steps. The ingot is delimited by the vertical, black dashed lines. Right panel: Flux of products at $y = 150$ for different aspect ratios $w/h$, as a function of the reaction parameter $p$.}
\end{figure}

Cross-sectional SEM images collected from the fracture surface of a np-Au sample after exposure to one short TMA/H$_2$O ALD cycle (1s/1s at 0.8 torr) are shown in Fig. \ref{Fig_4}. Deposition of Al$_2$O$_3$ is exclusively observed in a $\sim 20$~micron thick surface near layer that is separated by a sharp interface from uncoated (unreacted) material. In context of the present simulations, Al$_2$O$_3$ ALD represents the extreme case of a surface reaction with a reactive sticking probability $p_S^R$ of TMA of $10^{-3} - 10^{-4}$ and $0.00$ on empty and filled sites, respectively \cite{Ott}. \\
In contrast to our simulation, the product formed during TMA exposure (a chemisorbed $-0-$AlMe$_2$ species) does not desorb. The SEM cross-section shown in Fig. \ref{Fig_4}  reveals that a 1 s TMA/H$_2$O exposure (at 0.8 torr) is long enough to allow diffusion through the first 20 micron of the porous structure. The relatively thick reaction layer is a consequence of the low sticking coefficient and that the product does not desorb; desorption of the product would regenerate the surface near empty catalytic centres, leading to faster consumption of the reactant. Thus both the reactant sticking probability and lifetime of the product seem to determine how much the bulk of the sample contributes to product formation.
 The flux of products exhibits a steep rise in the first layers followed by a much slower increase inside the ingot is discussed in more detail below (see Fig. \ref{Fig_5}(a)). We suggest that the significantly higher conversion at the gas-facing side (reflecting different local gas-surface chemistry) is the underlying reason for the experimentally observed dependence of the reaction-induced coarsening kinetics on the sample-gas stream orientation, with substantial coarsening of the rear surface while the gas-facing side appears to be unaltered (Fig. \ref{Fig_1}).
 
Motivated by this result, we systematically studied the effect of sample thickness on conversion efficiency, $\eta$, with a focus on low reaction probabilities. The overall conversion efficiency is determined by the integral over reactant flux through the pores, the exposed surface area, and the reaction probability. Because the probability for transmission of a reactant molecule decreases linearly with increasing sample thickness, \cite{Gordon}, the number of molecules that actually diffuses far enough to take advantage of the additional internal surface area of thicker samples becomes very small, so that for low reaction probabilities the loss of unreacted reactants via backscattering through the surface facing the gas flux becomes the dominant loss mechanism for thicker samples. 

For a more quantitative analysis, we studied the effect of the relative time scales of mass transport and diffusion, and of the reaction probability on the conversion efficiency. The conversion efficiency is controlled by the ratio between the chemical reaction time scale $\tau_{ch} =\dt/p =1/p$ in lattice units, and the mass transport time scale $\tau_{tr}= \delta / \bar{v}$, where $\bar{v}$ is the average molecule velocity inside the pore, which we take of the order of the lattice speed of sound $c_s$;  this ratio is the Damk\"ohler number defined in Eq.~(\ref{Da_def}).\\
The limit $\Da \rightarrow 0$ corresponds to infinitely fast chemical reactions, in which case the reactivity is limited by transport phenomena, that is, how much reactant is brought in contact with the catalyst. In the opposite limit of infinitely slow chemistry, 
$ \Da \rightarrow \infty$, reactivity is limited by the timescale of the reactions. In the former limit, the thickness of the slab is not expected to play any major role, while in the latter it surely does because the reactant molecules must reside within the ingot sufficiently long to give molecules enough time to react. In our simulations $\Da  \sim 0.4$, marginally in the fast-chemistry limit. \\
In physical units, we estimate that $\tau_{tr}=50$ nm / 1000 m/s  $\sim$ 50 ps and $\tau_{ch}  \sim 25$ ps, as explained in earlier section. We define the conversion efficiency as the mass of the products at the outlet over the mass of reactant at the inlet, namely:
\be 
\eta =\frac{M_P |_{y=L}}{M_R |_{y=0} } \ ,  
\label{eff}
\ee
where $M_R (y)=\sum_x \rho_R(x,y)$  is the total mass of reactant at $\bm{x}$.
We studied the effect of ingot thickness $w$ (ranging from 6, 10, 20 to 30, all in lattice units with $\Delta x = 10$ nm) on the conversion efficiency $\eta$ while keeping the ratio between the ingot length h and the inner glass tube diameter $H$ constant ($200 : 500$), similar to the experimental value $1:2.5$. In all simulations, the porosity is fixed at  $\phi \sim  0.70  \pm 0.005$. 
\begin{figure}
\centering
	\includegraphics[width=0.46\textwidth]{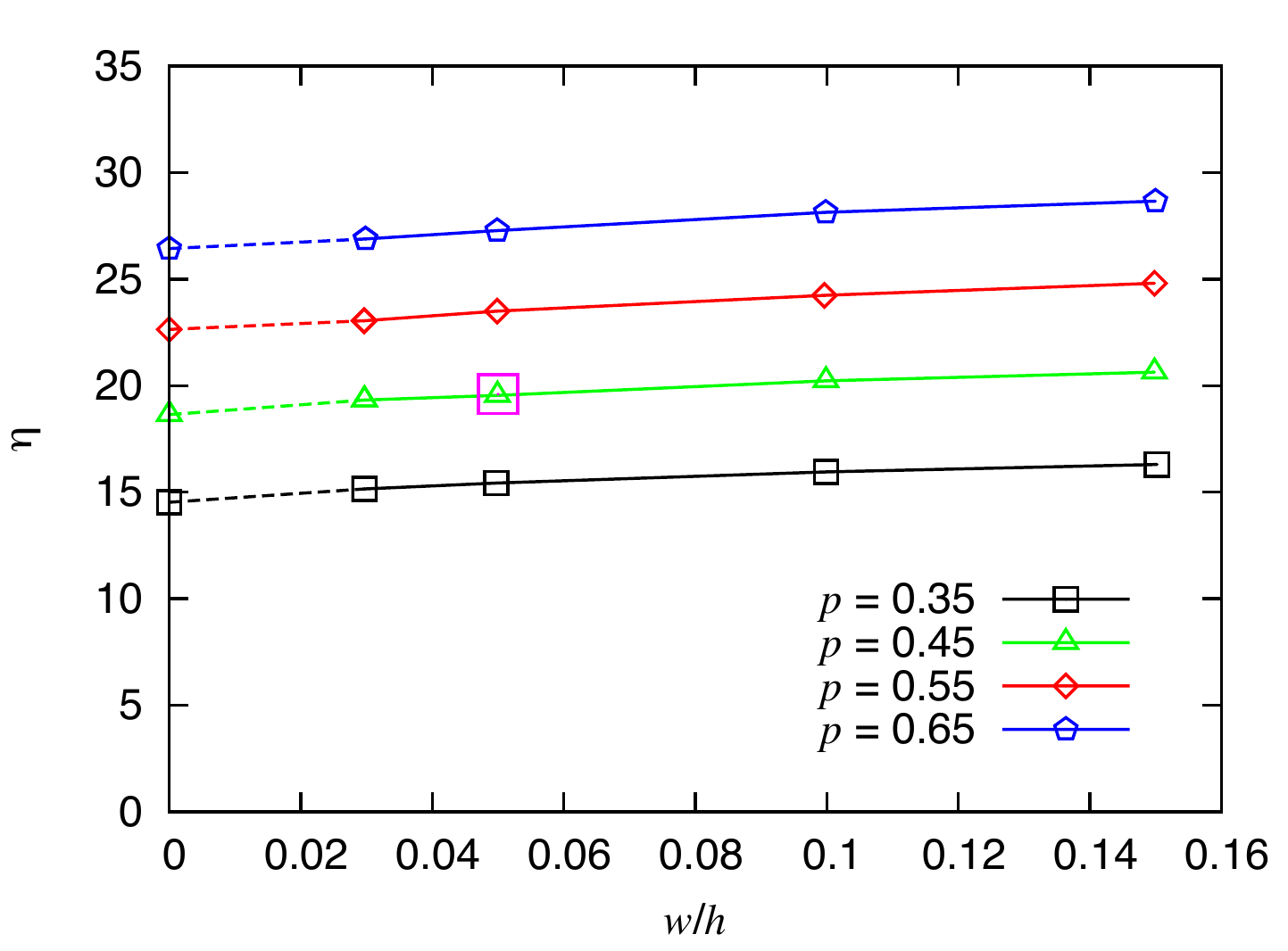}
	\caption{\label{Fig_6}: Trends in the efficiency $\eta$ in a horizontal ingot, as a function of w/h ratios, for different values of $p$ (pink open square denotes the experimental value), for a $1:2.5$ ratio between the ingot diameter (h) and the glass tube dimension ($H$). The fact that the trends exhibit a non-zero value of $\eta$ as $w/h \rightarrow 0$ indicates that the ingot perimeter (equivalent to surface area in 3D) plays a dominant role in the conversion process, rather than its area (volume in 3D).}
\end{figure}

Figure \ref{Fig_6} shows the trends of the conversion efficiency $\eta$ as a function of the parameters $w/h$ and $p$. In particular, we focused on the ratio $w/h = 0.05$ which corresponds to the actual ratio of the ingot used in the experiments (disk diameter 5 mm, disk thickness 250 $\mu$m). For $w = 10$ in lattice units, the experimentally measured conversion efficiency of $\eta=0.2$ is reproduced by choosing a reaction probability $p = 0.45$, corresponding to a characteristic chemical time of 25 ps. The trends highlight that the effect of the ingot thickness (i.e. its area in 2D and its volume in 3D) is negligible as compared to contribution attained by the ingot perimeter in the conversion process, as clearly visible by the values of $\eta$ for $w/h \rightarrow 0$. \\
The chemical efficiency in Eq. (\ref{eff}) can be related to the geometrical and chemo-physical parameters, as follows:
\be 
\eta = \frac{h}{H} \ \varphi \left( \frac{\bar{v} \ \tau_{ch}}{l} \right) \  ,  
\ee
where $l \sim w$ is the typical path length within the ingot, $\bar{v}$ the average molecular speed and $\tau_{ch}$ the chemical time scale. In the above expression, $\varphi \left( \frac{\bar{v} \ \tau_{ch}}{l} \right)$ denotes a generic (decreasing) functional dependence on the dimensionless quantity, $\Da=\bar{v} \ \tau_{ch}/l$, (Damk\"ohler number at the ingot level). This can also be decomposed as $\Da=(\bar{v} \ \tau_{ch}/ \delta) \times (\delta / l) \equiv \Da_p \times A_p$, where $\Da_p$ is the Damk\"ohler number at the pore level and $A_p= \delta/l$ is the aspect ratio of the pores.  \\
To study the effect of sample orientation relative to the gas flow, we also performed simulations for the case of an ingot placed parallel to the flow stream, as shown earlier in Fig. \ref{Fig_3}. This configuration is frequently used in filtration and, for the current application, may lead to higher efficiencies for the case of very high reaction probabilities that lead to very shallow product depth distributions. We have investigated the effect of the reaction parameter p on the reactant conversion, for a given aspect ratio, $w/h = 0.05$. We notice that even though the vertical setup exposes both faces of the plate, the sample now blocks a much smaller fraction of the gas flow so that most molecules bypass the ingot without ever coming in contact with it. Thus, despite the increase in exposed surface area, this leads to reduced conversion efficiency.
To investigate how thick a horizontal catalyst sample needs to be in order to still allow for efficient conversion, we performed a systematic study on the effect of the reaction probability $p$ (taking, $p = 0.35, 0.45, 0.55 \ \text{and} \ 0.65$) and the ingot dimension $w/h$ ($w/h = 0.03, 0.05, 0.1 \ \text{and} \ 0.15$), at the given porosity $ \phi \sim  0.70  \pm 0.005$. \\
The flux of products along the $y$ direction (i.e. perpendicular to the gas flow),
\be 
\Phi_P (y ) =  \sum_x  \ \rho_P (x,y) \ u(x,y) \ ,  
\ee
is shown in Fig. \ref{Fig_5}(b). This figure shows that for higher values of $p$, the depth of conversion remains almost unchanged and that the large part of reactant conversion takes place in the first few layers of the ingot, on the side facing the gas flow. \\
Figure \ref{Fig_5}, as well as Fig. \ref{Fig_6} and the density contour plots discussed earlier in Fig. \ref{Fig_3}, clearly show that most of the chemical conversion takes place very close to the outer surface of the porous catalyst sample that faces the inlet of the computational domain. For all values of $p$, the flux of products exhibits two trends across the ingot: a first, steep trend, at the ingot surface on the stream-facing side, and a second, slight slope for the flux of products through the ingot. The presence of these two trends support the conclusion of the major role played by the ingot perimeter in the conversion process, as highlighted in Fig. \ref{Fig_6}. At the outer surface, the trend of product flux shows no steep gradients, for all values of $p$: the difference in the trends at the two sides of the ingot is due to the different behaviour of the two faces in the conversion process, as confirmed by the experimental observations. The SEM inspection of the ingot sides, in fact, shows that the down-stream face is characterized by much more coarsened grains than the gas-facing one (see Fig. \ref{Fig_1}).

\section{Conclusions}
We have simulated chemical and transport phenomena in nano-sized pores of a catalytic ingot, by means of the Lattice Boltzmann Method. In all simulations, most of the reactions appear to take place at the front face of the ingot, with little or very minor involvement of the back  side. This is in good agreement with the experimental measurements and the SEM observations of the ingot surface, where different changes in the nano-structure appear to take place on the two faces of the ingot, according to the different contribution to chemical reactions.
From our simulations, we find that for the parameter regime investigated most of the reaction takes place in the first few layers of the ingot. The width $w$ of the ingot slightly increases the overall conversion efficiency, under the condition that the traversal time $w/ \bar{v}$ becomes comparable with the chemical time scale $\tau_{ch}$, i.e, the small Damk\"ohler number limit. We also explored the effect of the sample orientation relative to the flow direction, by posing the ingot long axis parallel to the flow. This yields a considerably lower conversion efficiency, due to aerodynamic effects which hinder the contact of the reactant molecules with the reactive surface. \\
The present code performs at $\sim 1.5$ MLUPS per second: namely, it updates $\sim 1.5$ million lattice sites per second, which is in line with the expected LB performance, taken into account that the code evolves three chemical species. \\
Future plans include shape optimization of single and multi-ingot configurations, as well as multiscale procedures to link the present mesoscale simulations to microscale and atomic models.

\begin{acknowledgements}

This work is supported by the Integrated Mesoscale Architectures for Sustainable Catalysis (IMASC) Energy Frontier Research Center (EFRC) of the Department of Energy, Basic Energy Sciences, Award DE-SC0012573. Work at LLNL was performed under the auspices of the U.S. Department of Energy by LLNL under Contract DE-AC52-07NA27344. We thank the Research Computing group of the Faculty of Arts and Sciences, Harvard University, for computational resources and support. \\
We thank Prof. Elio Jannelli, G. Di Staso and members of the IMASC EFRC, C.M. Friend, R.J. Madix, M. Flytzani-Stephanopoulos, for many valuable discussions. \\
C.B. acknowledges postdoctoral fellowships through the Belgian American Educational Foundation (BAEF) as well as Wallonie-Bruxelles International (Excellence grant WBI.WORLD) foundations. \\

\end{acknowledgements}




\section{Appendix}

\subsection{Numerical Method}

\begin{figure}
\begin{center}
	\subfigure[]{\includegraphics[width=0.25\textwidth]{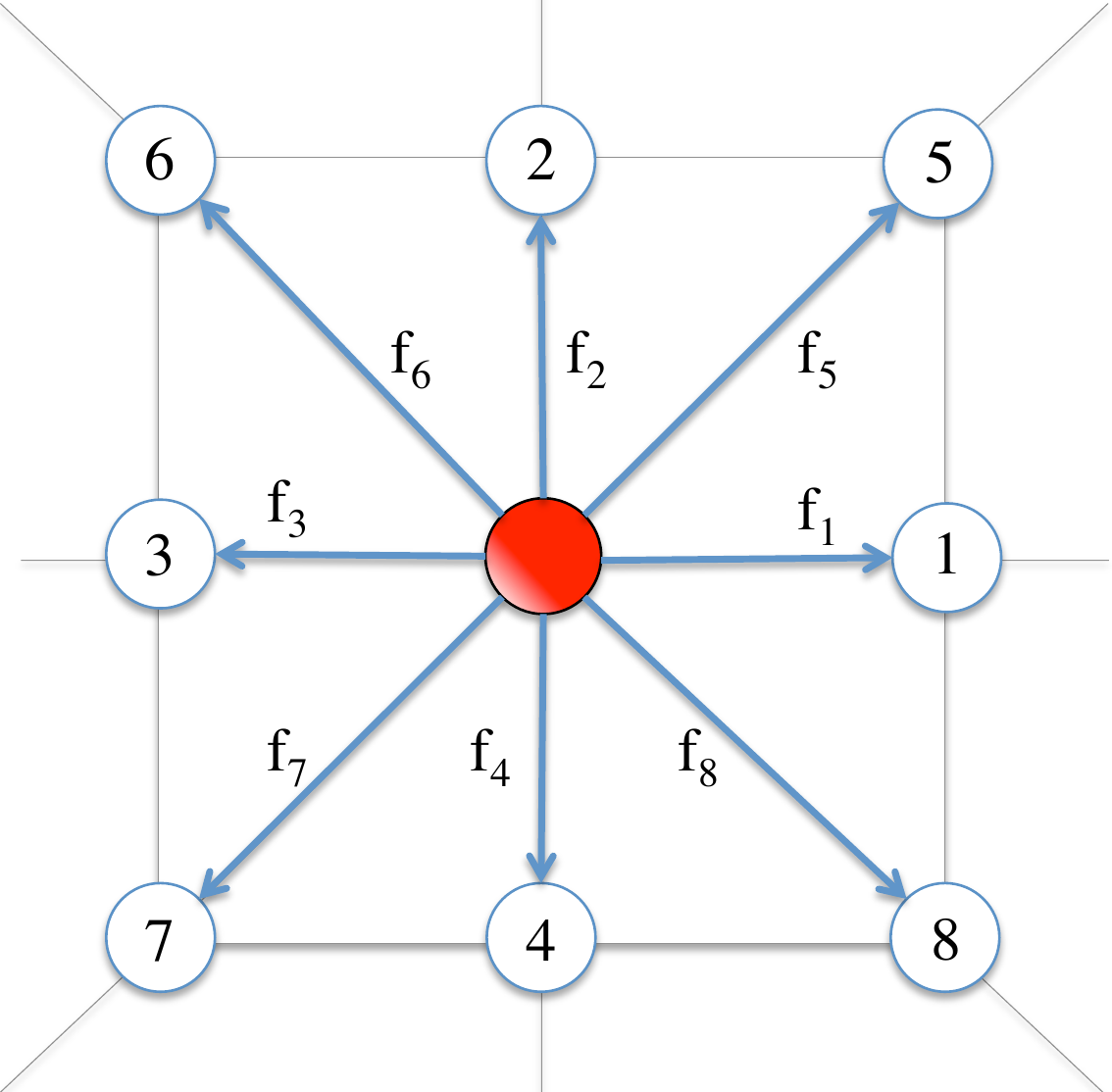}} \;
	\subfigure[]{\includegraphics[width=0.25\textwidth]{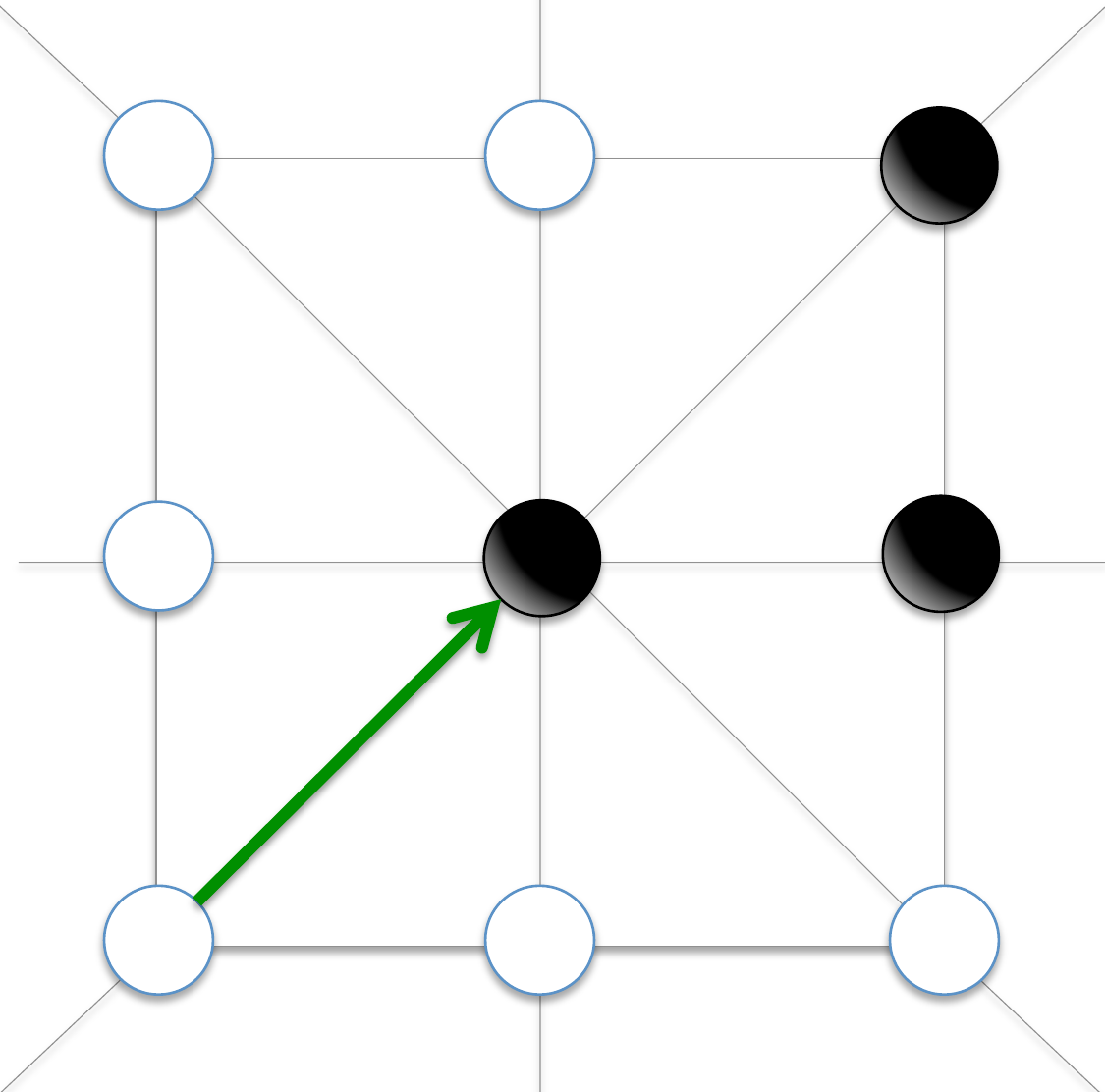}} \;
	\subfigure[]{\includegraphics[width=0.25\textwidth]{Figure_Ab.pdf}} 
\end{center}
\caption{\small{\label{Fig_A1}(a) Schematic representation of the nine-velocity lattice for 2D simulations with the four first (1- 4) and four second (5-8) nearest neighbour sites and corresponding labelled velocities. (b) and (c): Representation of the reactive boundary condition: the green arrow in (b) represents an  outgoing reactant molecule impinging on a solid site (black circle) from the pore bulk. The length of the arrow indicates the magnitude of the corresponding Boltzmann’s discrete population; in (c), the green (reactant) and red (product) arrows show the molecular distribution after the desorbing event. }}
\end{figure}

The three main features of the LBM are the following:
\begin{enumerate}
\item The distribution function is expanded onto a small basis set, given by a few Hermite polynomials. The kinetic moments of the distributions defined the macroscopic quantities of interest, i.e. density and momentum. Special nodes in velocity space are used to sample the hydrodynamic variables and collisional terms, as shown schematically in Fig. \ref{Fig_A1} for a two-dimensional (2D) case. This formulation dramatically simplifies the numerical task, since the distribution function is then turned into a small set of populations. Resorting to a small set of quadrature points is justified by the fact that, under standard fluid dynamics conditions, the distribution functions exhibit of microscopic exhibit a mild departure from the Maxwell-Boltzmann equilibrium;
\item The support in space is discretized over a cartesian mesh which represents a substantial simplification as compared to conventional numerical solvers of the continuum fluid dynamics equations. On the cartesian mesh, the fluid populations hop from one lattice site to a neighbouring one in a single time step, which gives rise to a very compact and efficient stream-collide algorithm. 
\item The method naturally allows the possibility to host additional physical effects, such as multi-component chemical reactivity and phase transition, at a minor implementation cost. 
\end{enumerate}
The simulations are extremely well suited to parallel computing, which enables large scale applications. \\
We wish to point out that to full-scale simulation of microreactors at near nanometric resolution would require extreme computational resources, which cannot be deployed on a routine basis.  A few numbers will help convey the point. A LBM simulation with ten billion sites, running over ten million timesteps can simulate the operation of a millimetric reactor over a time span of a few seconds at a base resolution of about 100 nm in space and 100 ns in time. Such full-scale simulation would require about $10^{20}$ floating-point operations (hundred Exaflops), thus taking about three years elapsed time on a Teraflop ($10^{12}$ Flops/s) computer. Hence, in order to bring this down to routine operation, say a few days elapsed time per simulation, a Petaflop ($10^{15}$ Flops/s) computer is needed. Thanks to extreme amenability of LBM to massively parallel implementation, such cutting-edge simulations have already appeared in the supercomputing literature, \cite{Bernaschi_2013}. At present, widely available computers fall far short of such performance on a routine basis, making it necessary to resort to simulating systems far smaller than the experimental ones.

In this paper, we employ the standard version of the LBM for a multicomponent system, with three species labelled by the index α:  an inert carrier ($C$), a reactant ($R$) and a product ($P$), in order to model the following methanol oxidation reaction:

\begin{equation}
2 \text{CH}_3 \text{OH} + \text{O}_2 \rightarrow \text{HCOOCH}_3  + 2 \text{H}_2 \text{O},
\end{equation}

The fluid-dynamic evolution is governed by the following discretized formulation of the Boltzmann kinetic equation, 
\be
f_i^{\alpha} (\x+\bm{c}_i, t+1)-f_i^{\alpha} (\x,t)=\omega [f_i^{eq,\alpha} (\x,t)-f_i^{\alpha} (\x,t)] \,  ,         
\label{LBE}
\ee
 
where $f_i^{\alpha} (\bm{x},t)$ is the probability density function of finding a particle at site $\bm{x}$ at time $t$, moving along the $i$-th lattice direction defined by the discrete speeds $\bm{c}_i$, with $i = 0,…,b$. The lattice time step $\dt$ has been taken as the unit of time for simplicity. The left hand-side of Eq. (2) represents the free-streaming of molecules, whereas the right-hand side accounts for the collisional relaxation towards the local equilibrium $f_i^{\alpha} (\x,t)$, which is expressed as a low-Mach, second-order expansion of a local Maxwellian, namely:
{\setlength\arraycolsep{2pt}
\begin{eqnarray}
f_i^{eq,\alpha} (\x ,t)  & = & w_i \ \rho^{\alpha} (\x ,t)  \left[1+ \frac{\bm{c}_i \cdot \bm{u}(\x,t)}{c_s^2}  + \right.  \nonumber \\
 & + & \left. \frac{\left(\bm{c}_i \cdot \bm{u}(\x,t)\right)^2}{2\ c_s^4}  - \frac{|\bm{u}(\x,t)|^2}{2 \ c_s^2} \right] \, .
\end{eqnarray}}
The relaxation to local equilibrium takes place on a time-scale $\tau=1/\omega$, taken to be equal for all species. \\
The first two moments of the distribution functions provide the macroscopic gas densities, $\rho^{\alpha} (\x,t)$, and velocities, 
$\bm{u}^{\alpha} (\x,t)$, respectively: 
\be 
\rho^{\alpha} (\x,t) = \sum_{i=0}^b f_i^{\alpha} (\x,t) \ , \quad 
\rho^{\alpha} (\x,t) \bm{u}^{\alpha} (\x,t)= \sum_{i=0}^b \bm{c}_i \ f_i^{\alpha} (\x,t)\ .
\ee
At local equilibrium all species move with the common barycentric velocity:
\be 
\bm{u}(\x,t)= \sum_{\alpha} \rho^{\alpha}(\x,t) \ \bm{u}^{\alpha}(\x,t) / \sum_{\alpha} \rho^{\alpha}(\x,t)
\ee
The weights $w_i$ in the equilibrium distribution satisfy the following conservation and isotropy constraints:
\be 
\sum_{i=0}^b w_i=1 \ , \quad \sum_{i=0}^b w_i c_i=0 \ , \quad \sum_{i=0}^b w_i c_i c_i=c_s^2 \ \bm{1} \ ,
\ee
where $\bm{1}$ denotes the unit matrix. In our implementation we consider a nine discrete speed ($b = 8$) scheme, also called the D2Q9 lattice. The speeds are $c_i = (0; 0)$, corresponding to a weight $w_i  = 4/9$, $c_i  = (\pm 1; 0)$ and $c_i  = (0; \pm 1)$, with  $w_i= 1/9$, and $c_i=(\pm 1;\pm 1)$, with $w_i  = 1/36$. Finally, $c_s$ is the lattice sound speed equal to $1/\sqrt{3}$. \\
It can be shown that, in the limit of small Knudsen number (defined as the ratio of mean free path to macroscopic length scale), Eq~(\ref{LBE}) reproduces the Navier-Stokes equation for a carrier fluid of viscosity $\nu =c_s^2 (\tau -1/2)$; in the hydrodynamic regime, 
$\tau - 1/2 << 1$, in lattice units. In the present application, which deals with gas flows characterized by Knudsen numbers $\Kn \sim 1$, we work in the regime $\tau - 1/2 \sim 1$. The remaining two LB equations reproduce the dynamics of the reactants and products as advected by the carrier and diffuse across it.

Species interconversion due to catalytic reactions at the pore surface is accounted for by considering a local exchange of populations, as they meet and react on the solid walls of the pore. In the case of heterogeneous catalysis, such operation takes places when gas populations hit the surface of the porous catalyst. Here we consider a first order chemical reaction of the generic form 
$R \rightarrow P$, whereby the interaction of the reactant with the porous medium is assumed to be an instantaneous interconversion into the product upon contact with the porous catalytic walls (no reaction takes place in the gas region). The assumption of instantaneous reaction relies on a clearcut separation of timescales between the molecular collisions in the bulk and the chemical reactivity at the surface. 

We have implemented a ``sputtering'' boundary condition, as depicted in Figs. \ref{Fig_2} and \ref{Fig_A1}.  Specifically, the populations of reactants and products that enter or exit (superscripts ``in'' and ``out'') a node obey the following equations:
\bea
f_i^{R,in} (\x) &=& (1-p_S^R ) \  \sum_{j=1}^{b(\x)} S_{i,j} \  f_j^{R,out} (\x-\bm{c}_j ) \nonumber  \\ 
f_i^{P,in} (\x)  &=& (p \ p_S^R) \ \sum_{j=1}^{b(\x)} S_{i,j} \ f_j^{R,out} (\x-\bm{c}_j )  \nonumber \\
&+& (1-p_S^P) \sum_{j=1}^{b(\x)} S_{i,j} f_j^{P,out} (\x-\bm{c}_j ) \nonumber \ ,  \\
{} & & {} 
\label{f_P_in}  
\eea
$S_{i,j}$ is a random sputtering matrix, expressing the probability of a molecule leaving the bulk along direction ``$j$'',  to re-enter along direction ``$i$''. Since sticking and reaction events are already accounted for by the coefficients $p_S^R$, $p_S^P$ and $p$, the sputtering matrix obeys the conservation rule $\sum_{i=1}^{b(\x)} S_{i,j}=1$.

Note that $0 \leq b(\bm{x}) \leq 8$ is the number of active links at any given lattice site $\bm{x}$. Each node of the porous medium in direct contact with the gas nodes acts as a catalytic node.  The interpretation of the above equations is transparent: the right hand side of Eq. (\ref{f_R_in}) is the scattering back to the pore of the reactant molecules which did not stick to the wall, whence the prefactor $(1-p_S^R )$. The first term at the right hand side of Eq. (\ref{f_P_in}) represents the fraction of product molecules entering the bulk after reacting with reactant molecules stuck to the wall, whence the prefactor $p \ p_S^R$. The second term represents the product molecules which enter a solid node from the bulk and are re-emitted with probability $(1-p_S^P)$ without remaining stuck to the wall. Clearly, the flux of products from the solid to the fluid results from the sum of these reactive and non-reactive events. Note that the above boundary condition does not preserve the total $(R+P)$ mass unless both sticking probabilities are set to zero. The random, although discrete, orientation of the incoming populations mimics the mesoscopic effect of the wall orientation on the gas dynamics and the fact that the particles are adsorbed and released by the catalyst on a sufficiently long timescale to randomize the directions of the incoming particles.

\subsection{Validation of Numerical Model}

\begin{figure}
\begin{center}
	\includegraphics[width=0.4\textwidth]{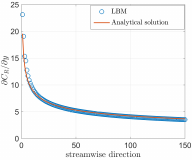}
\end{center}
\caption{\small{\label{Fig_A2} Comparison between analytical solution of the 2D laminar reactive flow between parallel plates. The concentration gradient along the crossflow direction at the reaction plate is plotted against the streamwise direction.}}
\end{figure}

We validate the novel chemical boundary condition against L\'ev\^eque's analytical solution \cite{Leveque} of a 2D laminar reactive flow between parallel plates. Reactions take place at the bottom plate. The vertical gradient of concentration at the reactive wall can be calculated as:
\begin{equation}
\frac{\partial [R]}{\partial y} = \frac{1}{9^{\frac{1}{3}}  \ \Gamma \left(\frac{4}{3}\right)} \ \left(\frac{4 \mathrm{Pe} }{\frac{x}{L_y}} \right)^{\frac{1}{3}}
\end{equation}
where $[R]$  is the reactant concentration, $\Gamma$ is the Gamma function, $x$ is the streamwise direction, $\mathrm{Pe}$ is the Pecl\'et number, set to $\sim 253$ for the simulation at hand, and $L_y$ is the number of nodes along the crossflow direction $y$. The benchmark test was carried out on a $150 \times 50$ grid. A body force has been used to simulate the constant pressure gradient along the channel. The reactant is injected continuously at the inlet of the channel reacting only with the lower plates. The reaction probability in the sputter boundary conditions has been set to 1, so as to simulate instantaneous reaction ($Da_A \rightarrow \infty$), i.e. all the reactant impacting on the active nodes are instantly converted into product).
Good agreement with the analytical solution is found, as evidenced by the comparison reported in Fig. \ref{Fig_A2}.
The overall error, measured as a Mean Absolute Error, is roughly $4\%$.

\end{document}